\documentclass[onecolumn,showpacs,preprintnumbers,amsmath,amssymb]{revtex4}
\usepackage{graphicx}
\usepackage{dcolumn}
\usepackage{bm}
\usepackage{epsfig}
\usepackage{subfigure}

\newcommand{\bi}{\begin{itemize}}
\newcommand{\ei}{\end{itemize}}
\newcommand{\be}{\begin{eqnarray}}
\newcommand{\ee}{\end{eqnarray}}
\newcommand{\beq}{\begin{equation}}
\newcommand{\eeq}{\end{equation}}
\newcommand{\bbmatrix}{\left( \begin{array}}
\newcommand{\eematrix}{\end{array} \right)}
\newcommand{\bbbmatrix}{\left[ \begin{array}}
\newcommand{\eebmatrix}{\end{array} \right]}

\begin{document}

\title{Electron Correlation in Oxygen Vacancy in SrTiO$_3$}

\author{Chungwei Lin and Alexander A. Demkov\footnote{E-mail: demkov@physics.utexas.edu}}
\date{\today}
\affiliation{Department of Physics, University of Texas at Austin, Austin, Texas 78712, USA}

\begin{abstract}
Oxygen vacancies are an important type of defect in transition metal oxides. 
In SrTiO$_3$ they are believed to be the main donors in an otherwise intrinsic crystal. 
At the same time, a relatively deep gap state associated with the vacancy is widely reported.  
To explain this inconsistency we investigate the effect of electron correlation in an 
oxygen vacancy (OV) in SrTiO$_3$. When taking correlation into account, 
we find that the OV-induced localized level can at most trap one electron, 
while the second electron occupies the conduction band. 
Our results offer a natural explanation of how the OV in SrTiO$_3$ can produce 
a deep in-gap level (about 1 eV below the conduction band bottom) in photoemission, 
and at the same time be an electron donor. 
Our analysis implies an OV in SrTiO$_3$ should be fundamentally regarded as a magnetic impurity,
whose deep level is always partially occupied due to the strong Coulomb repulsion.
An OV-based Anderson impurity model is derived, and its implications are discussed.
\end{abstract}

\maketitle


Perovskite SrTiO$_3$ is a band insulator with a band gap of 3.2 eV \cite{J.Appl.Phys.90.6156,Zollner},
which undergoes a structural phase transition around 110 K \cite{PhysRev.177.858}, and displays 
quantum paraelectricity (a very large but finite static dielectric constant) below 4 K \cite{PhysRevB.19.3593}.
Many truly fascinating phenomena  
such as superconductivity \cite{PhysRevLett.12.474, PhysRev.163.380}, 
ferromagnetism \cite{Brinkman_07, Li_11, Bert_11}, Kondo resistance minimum \cite{PhysRevLett.107.256601},
and two-dimensional electron gas \cite{Ohtomo_04,Thiel_06,Santander-Syro_11,NatMetal_STO_2D} occur,
when a small number of conduction electrons are introduced into SrTiO$_3$.
The electrons can be provided by doping SrTiO$_3$ on either the A (i.e. Sr$_{1-x}$La$_x$TiO$_3$  \cite{PhysRevB.61.12860})
or B (i.e. SrTi$_{1-x}$Nb$_x$O$_3$ \cite{PhysRevLett.100.056401}) sublattices, or by
introducing impurities or defects such as hydrogen atoms \cite{PhysRevLett.108.116802, PhysRevB.79.035311, Hatch_13}
or oxygen vacancies (OV) \cite{Kan_05,NatMetal_STO_2D}. The OV in SrTiO$_3$ is particularly intriguing. On the one hand, 
the OV concentration is roughly proportional to that of the carriers \cite{Kan_05,PhysRevLett.98.196802,NatMetal_STO_2D}, 
strongly suggesting that vacancies are electron donors; on the other hand, an OV produces 
an in-gap signal, peaked approximately 1.0 eV below the conduction band, in angle resolved photoemission spectroscopy (ARPES),
even at temperatures as low as 20 K \cite{OV_Arpes_2002,NatMetal_STO_2D, Hatch_13}.
Within the single-particle description, 
the ARPES spectrum implies that an OV results in a deep impurity level and therefore is not likely to be the electron donor.
In this Letter, we demonstrate that, by taking the correlation effect, these two seemingly conflicting observations can be naturally reconciled.
More fundamentally, our analysis suggests that an OV in SrTiO$_3$ should be regarded as a magnetic impurity,
in the sense that a deep level is always partially occupied due to the strong Coulomb repulsion, and may account for the observed Kondo physics \cite{PhysRevLett.107.256601} 
or interface ferromagnetism \cite{Brinkman_07, Li_11, Bert_11}.

\begin{figure}[http]
\epsfig{file=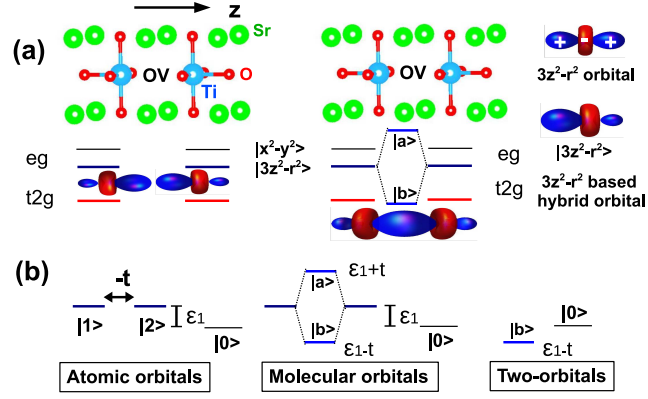, width = 0.5\textwidth}
\caption{(Color online) (a) Main changes in electronic structure due to an OV as described in Ref. \cite{PhysRevB.86.161102}.
(Left) Due to the local C$_{4v}$ symmetry, the $3d_{3z^2-r^2}$ of the Ti atom next to an OV hybridizes with the Ti $4p_z$ orbital, forming 
a $3d_{3z^2-r^2}$-based hybrid state, whose energy is lower than that of Ti $3d_{x^2-y^2}$ level.
$|3z^2-r^2 \rangle$ represents $3d_{3z^2-r^2}$-based hybrid state that has an asymmetric profile due to the non-negligible Ti $4p_z$ component.
(Middle) Two Ti $3d_{3z^2-r^2}$-based hybrid orbitals further hybridize to form the bonding $|b \rangle$ and
anti-bonding $|a\rangle$ states. (Right) The difference between the unperturbed $3d_{3z^2-r^2}$ orbital and the
$4p_z$ mixed $3d_{3z^2-r^2}$-based hybrid orbital is illustrated. Blue (+) and red (-) colors indicate the sign of the wave function.
(b) A simplified three-orbital model in the atomic (left) and molecular (middle) orbital pictures.
The on-site repulsion only applies to $|1 \rangle$ and $|2 \rangle$, which represent two Ti $3d_{3z^2-r^2}$-based hybrid orbitals, due to
their localized nature. $|0 \rangle$ represents a delocalized bath level (of Bloch form).
(Right) The simplest model keeping only two low-energy orbitals -- one correlated orbital $|b\rangle$ and one bath orbital $|0\rangle$. 
}
\label{fig:illustration}
\end{figure}

The electronic structure of an OV in SrTiO$_3$ has been extensively studied using density functional theory (DFT) 
\cite{PhysRev.140.A1133, Robertson_2003, PhysRevB.68.224105,PhysRevB.70.214109,  Hou_10, PhysRevB.86.155105, PhysRevB.86.195119,  PhysRevB.73.064106}.
Depending on the functional used, the OV induced impurity level in SrTiO$_3$ can lie above the
conduction band edge, thus leading to a resonance \cite{PhysRevB.70.214109, PhysRevB.86.155105, PhysRevB.86.195119};  
be an in-gap bound state with energy level position ranging from 0.4 to 1 eV below the conduction band bottom \cite{PhysRevB.68.224105, PhysRevB.73.064106,  PhysRevB.86.155105};
or be a partially filled, spin-polarized gap state \cite{Hou_10}.
{\it Independent} of the functional, the OV-induced state is spatially localized 
and is mainly composed of the next-to-OV Ti $3d_{3z^2-r^2}$ and $4p_z$ orbitals \cite{PhysRevB.86.161102}.
To include the correlation effect in an OV, we first identify the main change in the local electronic structure that it causes. The following discussion is based on Ref. \cite{PhysRevB.86.161102}. 
Under cubic symmetry, the proper local orbital basis of a Ti atom consists of two $3d_{x^2-y^2}$,  $3d_{3z^2-r^2}$ orbitals with $e_g$ symmetry,
and three $3d_{xy}$, $3d_{yz}$, $3d_{xz}$ orbitals of $t_{2g}$ symmetry. The vacancy mainly affects the
$3d_{3z^2-r^2}$ orbital, and its minor effect  on the other four $3d$ orbitals will be neglected.
Specifically, an OV reduces the local symmetry from cubic to C$_{4v}$, which allows for a local mixing 
between Ti $3d_{3z^2-r^2}$ and $4p_z$ ($z$ is along the Ti-OV-Ti direction), leading to a new $3d_{3z^2-r^2}$-based hybrid orbital.
Compared to the $3d_{3z^2-r^2}$ orbital without OV, the $3d_{3z^2-r^2}$-based hybrid is substantially lower (about 1.0 eV)
in energy and has a non-negligible Ti $4p_z$ component  \cite{caution1}, leading to an asymmetric shape
as illustrated in Fig.~\ref{fig:illustration}(a).
The more spatially extended  $4p_z$ component further allows for hybridization between 
the two   $3d_{3z^2-r^2}$-based hybrids adjacent to the vacancy. 
Their bonding combination gives the lowest OV-induced single-particle level
(middle panel of Fig.~\ref{fig:illustration}(a)).

The essential physics just described, is well captured by the following three-orbital model:
\beq
\begin{split}
H_{3o} &= 
\varepsilon_1 \sum_{\sigma} (n_{1,\sigma} + n_{2,\sigma}) -t \sum_{\sigma} [ c^{\dagger}_{1,\sigma} c_{2,\sigma}  + h.c. ] \\
&+ U_1 \sum_{i=1,2} n_{i,\uparrow} n_{i,\downarrow} + \varepsilon_0 \sum_{\sigma} n_{0,\sigma}
\end{split}
\label{eqn:H3o}
\eeq
with $\sigma$ labeling spin. Here 1 and 2 represent two Ti $3d_{3z^2-r^2}$-based orbitals of energy $\varepsilon_1$, 
and their spatial overlap is described by a hopping parameter $t$. 0 represents
the uncorrelated (bath) orbital of energy $\varepsilon_0$ which is set to zero (Fig.~\ref{fig:illustration}(b), left panel).
Because the bath orbital is of Bloch form which extends throughout the entire sample, its correlation can be neglected \cite{bath_orbital}.
The on-site repulsion $U_1$ is applied only to two spatially localized orbitals 1 and 2. 
Equivalently, to describe the same Hamiltonian, one could  use molecular orbitals 
$|b \rangle = [|1\rangle + |2\rangle]/\sqrt{2}$,  $|a \rangle = [|1\rangle -|2\rangle]/\sqrt{2}$
('b', 'a' stand for bonding and anti-bonding respectively),
that diagonalize the quadratic part of Eq.~\eqref{eqn:H3o}
(Fig.~\ref{fig:illustration}(b), middle panel).
Since each OV provides two electrons, one has to determine the two-electron ground state
of Eq.~\eqref{eqn:H3o}. Depending on the parameters of the Hamiltonian, the ground state falls into one of three categories
-- Type I with both electrons occupying the bath orbital (with no electrons occupying the
OV-induced state), Type II with one electron occupying the bath orbital (one electron occupying the OV-induced state), or  
Type III, when neither electron occupies the bath orbital (both electrons occupying the OV-induced state). Their respective energies are 
\beq
\begin{split}
E_{I} &= 0\\
E_{II} &= \varepsilon_1 - t\\
E_{III} &= 2 \varepsilon_1 + \frac{1}{2} \left(U_1 - \sqrt{U_1+16t^2} \right).
\end{split}
\label{eqn:energies}
\eeq

\begin{figure}[http]
\epsfig{file=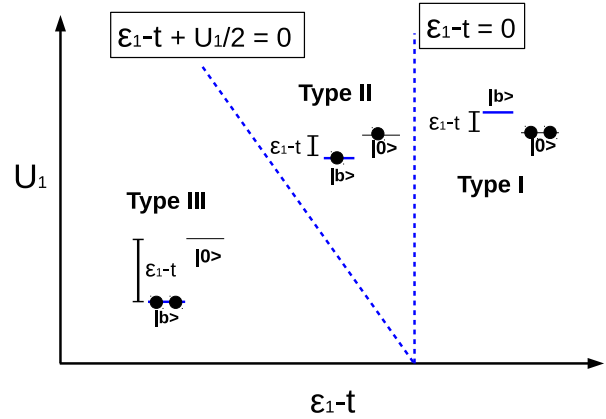, width = 0.5\textwidth}
\caption{(Color online) The phase diagram of $H_{3o}$ (using the small $U_1$ expansion) in Eq.~\eqref{eqn:H3o} as a function of $\varepsilon_1-t$ and $U_1$. 
Two phase boundaries are given by $\varepsilon_1-t = 0$ and $\varepsilon_1-t +U_1 /2=0$.
The Type I state corresponds to DFT using LDA, whereas the Type III to using a hybrid functional.
The Type II state corresponds to DFT using a spin-polarized GGA+U functional in terms of orbital occupation, 
but does not have any magnetic moment.
The Type II state is the best description of the OV behavior in SrTiO$_3$.
The electron distributions of three types of ground state are indicated. 
}
\label{fig:phase_diagram}
\end{figure}

When $\varepsilon_1 - t > 0$, the ground state is of Type I (both electrons occupying the bath level) regardless of the value of $U_1$.
This corresponds to a DFT calculation in the local density approximation (LDA) 
\cite{PhysRevB.70.214109, PhysRevB.86.155105, PhysRevB.86.195119}.
When $\varepsilon_1 - t < 0$, the ground state can be of Type II or III depending on the value of $U_1$.
In the large $U_1$ limit, the ground state is of Type II as $E_{III} \sim 2 \varepsilon_1 - 4t^2/U_1 > 0 > E_{II}$.
The ground state is of Type III when $E_{III} < E_{II}$. 
Expanding $E_{III}$ in the small $U_1$ limit,  this condition leads to $\varepsilon_1 -t + U_1/2<0$.
The phase diagram in the small $U_1$ limit, depends only on $\varepsilon_1 - t$ and $U_1$ (rather than
$\varepsilon_1$, $t$, $U$), and is shown in Fig.~\ref{fig:phase_diagram}.
From the hybrid functional calculations \cite{PhysRevB.86.155105,HSE}, we estimate that $t \sim 1.2$ eV and $\varepsilon_1 \sim 0.4$ eV leading to an $\varepsilon_1 -t$ difference of -0.8 eV.
The energy cost of doubly occupying one Ti $3d$ orbital $U_1$ has been estimated to range between 
3.0 and 5.0 eV \cite{PhysRevB.74.054412}.
With these parameters, the ground state of a vacancy is of Type II, where the OV-induced state only traps one electron.
The Type II state is energetically favored when doubly occupying the bonding state costs too much energy.
We can further simplify the model by considering only the bonding orbital $|b \rangle$ and bath orbital $| 0 \rangle$ (Fig.~\ref{fig:illustration}(b), right panel). The effective two-orbital model of Eq.~\eqref{eqn:H3o} is
\beq
\begin{split}
H_{2o} &=  \varepsilon_0 \sum_{\sigma} n_{0,\sigma} +
(\varepsilon_1 -t) \sum_{\sigma} n_{b,\sigma} 
+ U  n_{b,\uparrow} n_{b,\downarrow}.
\end{split}
\label{eqn:H2o}
\eeq
In this orbital basis, doubly occupying the bonding orbital
costs energy $U = U_1/2$ \cite{molecular_U}. Note that $H_{2o}$ in Eq.~\eqref{eqn:H2o} gives 
the same phase diagram as that shown in Fig.~\eqref{fig:phase_diagram}, and can therefore 
be used as an effective model if only the bath occupancy is of interest. 
Also note that, when a hopping between the bath and bonding orbitals ($-g \sum_{\sigma} c^{\dagger}_{0, \sigma} c_{b, \sigma} + h.c.$, whose origin will be discussed later) is allowed,
the ground state is a spin-$singlet$ without a net magnetic moment \cite{Grosso}. 
We believe that the Type II state describes an OV in SrTiO$_3$, which should be fundamentally regarded as a magnetic impurity where 
the deep level can only be partially occupied.
As the Type II state has two partially filled single-particle orbitals, it is $not$ included in
the Hartree-Fock or mean-field type approach without symmetry breaking, because 
in this case each orbital is either empty or doubly filled. 
The Type II state is indeed found in a spin-polarized DFT calculation using the
generalized gradient approximation with a Hubbard U (GGA+U) \cite{Hou_10}, at the price of having a net magnetic moment.

\begin{figure}[http]
\epsfig{file=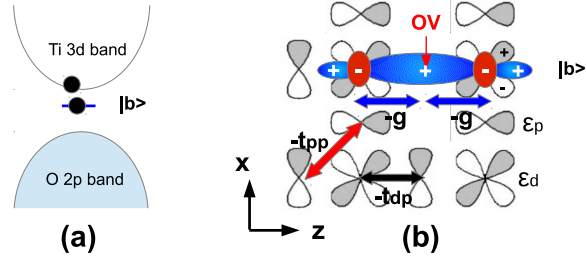, width = 0.5\textwidth}
\caption{(Color online) (a) The Anderson impurity model. The correlated OV-induced impurity orbital 
(the bonding orbital) which can only trap one electron.
(b) The bath orbitals are tight-binding bands composed of Ti $3d_{xz}$, O $2p_x$, O $2p_z$ orbitals 
(of energies $\varepsilon_d$, $\varepsilon_p$). 
The first and second  neighbor hoppings $t_{pd}$ and $t_{pp}$ are included.
 The OV-induced bonding state $|b \rangle$ can couple to two Ti $3d_{xz}$ orbitals across the OV. 
 Note that the coupling $g$ is zero when the local  C$_{4v}$ symmetry is preserved.
}
\label{fig:OV_AIM}
\end{figure}

By keeping only the bonding orbital $| b \rangle$, we introduce an OV-based 
Anderson impurity Hamiltonian \cite{PhysRev.124.41} $H_{imp} = H_{bath} + H_{OV}$. 
The bath part represents the $t_{2g}$ bulk bands described by the tight-binding approximation from Ref. \cite{PhysRevLett.97.056802}
\beq
\begin{split}
H_{bath} &= \varepsilon_d   \sum_{i} d^{\dagger}_i d_i +\varepsilon_p \sum_{i,o=x,z} p^{\dagger}_{i,o} p_{i,o} \\
&-t_{pd} \sum_i  [ d^{\dagger}_i p_{i,x} - d^{\dagger}_i p_{i-\hat{x},x} + d^{\dagger}_i p_{i,z} - d^{\dagger}_i p_{i-\hat{z},z} + h.c. ] \\
&-t_{pp} \sum_i  [ p^{\dagger}_{i,x} p_{i,z} - p^{\dagger}_{i,x} p_{i+\hat{x},z} - p^{\dagger}_{i,x} p_{i-\hat{z},z} 
+ p^{\dagger}_{i,x} p_{i+\hat{x}-\hat{z},z} + h.c. ],
\end{split}
\label{eqn:H_t2g}
\eeq
where $\varepsilon_d$,  $\varepsilon_p$ are energies of Ti $3d_{xz}$ (denoted by $d^{\dagger}_i$) and O $2p$   (denoted by $p^{\dagger}_{i,o}$) orbitals respectively, and
$t_{pd}$ is the hopping between the nearest neighbor Ti $3d_{xz}$ and O $2p$ orbitals, $t_{pp}$
is between two second neighbor O $2p$.
The correlated impurity orbital and its coupling to the bath are described by
\beq
\begin{split}
H_{OV} &=  \varepsilon_{imp}  \sum_{\sigma}   c^{\dagger}_{b,\sigma} c_{b,\sigma}  
+ E_{OV} p^{\dagger}_{i,z} p_{i,z}\\
&- g \left[ c^{\dagger}_{b,\sigma} d_{i,\sigma}  + c^{\dagger}_{b,\sigma} d_{i+\hat{z},\sigma}  + h.c. \right].
\end{split}
\label{eqn:H_imp}
\eeq
Here $c^{\dagger}_{b,\sigma}$ denotes the bonding orbital, while $g$ is the coupling between 
the  bonding (impurity) orbital and two Ti $3d_{xz}$ orbitals. The vacancy site is modeled by adding a large on-site potential $E_{OV}$ (we use 20 eV).
The model is schematically shown in Fig.~\ref{fig:OV_AIM}.
By fitting the ARPES bands \cite{PhysRevB.79.113103,Hatch_13}, we choose
$\varepsilon_d -\varepsilon_p = 4$ eV ($\varepsilon_p \equiv -0.8$ eV so that the valence band top is at zero energy), 
$t_{pd} = 1.5$ eV, and $t_{pp} = 0.2$ eV. This choice leads to the experimental band gap of 3.2 eV. We take $U=U_1/2 = 2$ eV
and $\varepsilon_{imp} = 2.4$ eV which is 0.8 eV below the conduction band bottom \cite{PhysRevB.86.155105}.

Using this model, we compute the zero-temperature ARPES spectrum $\rho_v(\omega)$ due to the impurity level by evaluating the  imaginary part of the hole Green's function \cite{PhysRevB.28.4315,Hufner}
\beq
\rho_v(\omega)  = \frac{1}{\pi} \text{Im} \left[ 
\langle  \phi_0 \left| c^{\dagger}_{b,\sigma} \frac{1}{\omega - i \eta -E_{GS} + H}  c_{b,\sigma} \right| \phi_0 \rangle \right],
\label{eqn:valence}
\eeq
where $| \phi_0 \rangle$ is the ground state of energy $E_{GS}$.
Numerically, we choose $\eta = 0.05$ eV. 
The matrix element in Eq.~\eqref{eqn:valence} is computed using the configuration interaction solver \cite{Helgaker, Sherrill, PhysRevB.86.165128, PhysRevB.88.035123}.
We keep 30 bath orbitals generated by the Lanczos procedure \cite{Grosso},
and the Fermi energy $E_F$ is chosen to be 3.4 eV, slightly above the conduction band bottom. 

\begin{figure}[http]
\epsfig{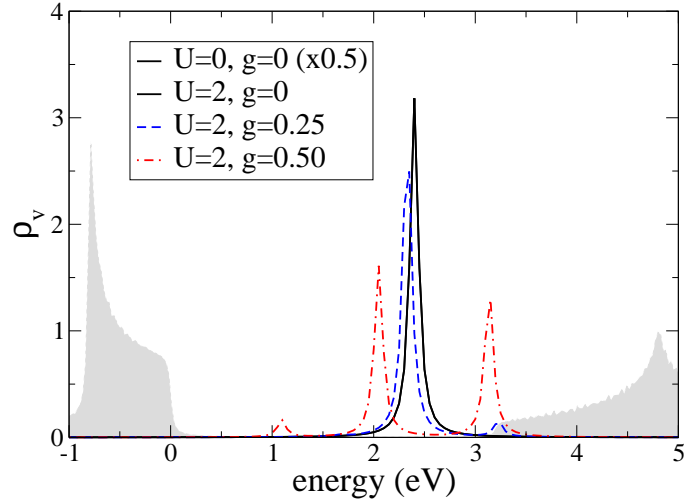}
\caption{(Color online) The spectral function $\rho_v$ for bath parameters $\varepsilon_d = 3.2$ eV,
$\varepsilon_p = -0.8$ eV, $t_{pd} = 1.5$ eV, and $t_{pp} = 0.2$ eV, which produce a band gap of 3.2 eV.
The bulk density of states is shown as the filled area for reference. 
The impurity level $\varepsilon_{imp}$ is 2.4 eV, 0.8 eV below the conduction band bottom.
Solid (black) curve: $U=0$, $g=0$ ($\times$ 0.5) and $U=2$ eV, $g=0$; 
dashed (blue) curve: $U=2$ eV, $g=0.25$ eV; dash-dotted (red) curve: $U=2$ eV, $g=0.5$ eV.
Once $U$ is larger than 1 eV, the in-gap state can have at most one electron (including spins).
The small satellite peak around 1.1 eV appears for $g=0.5$ eV.
}
\label{fig:spectra}
\end{figure}

We now discuss the impurity-bath coupling in the OV context.
We first note that due to its spatial extension, 
the impurity orbital can only overlap with two Ti $3d_{xz}$ orbitals adjacent to the OV
(the third term in Eq.~\eqref{eqn:H_imp}).
Moreover, the coupling is zero when the C$_{4v}$ symmetry is preserved,
as illustrated in the right panel of Fig. ~\ref{fig:OV_AIM}(b). 
In this case, the impurity is completely detached from the bath and
the problem can be trivially solved. In Fig.~\ref{fig:spectra} we see that the spectral functions
of $U=0$, $g=0$ and $U=2$ eV, $g=0$, are both peaked at $\varepsilon_{imp} = 2.4$ eV.
The weight (integrated spectral function over the peak) of the former is twice as large
as the latter.
This simply reflects that once $U$ is large enough, the impurity level can only trap one electron (including spin).
The coupling $g$ becomes nonzero when the local C$_{4v}$ symmetry is lifted, which can happen
at or near the interface, or under an applied electric field. With the nonzero impurity-bath coupling $g$,
the impurity state becomes less localized  \cite{PhysRevB.73.064106}.
In Fig.~\ref{fig:spectra} we show the vacancy spectral functions for $U=2.0$ eV and two values of the coupling constant $g=0.25$ and 0.5 eV. 
Three effects due to a nonzero $g$ can be seen. First, the main peak close to $\varepsilon_{imp}$
is pushed to a lower energy, which can be understood as the level repulsion between the conduction and
impurity orbitals. Second, some spectral weight near the conduction band appears due to
the impurity-bath hybridization. 
Lastly, a small satellite peak appears at lower energy  (at $\sim$1.1 eV) that originates from the
particle-hole pairs induced by the impurity-bath scattering.
We find this satellite peak to be quite general in the Anderson impurity model. 
A more detailed analysis will be provided in subsequent work.
For transport measurements, we note that the nonzero coupling maps $H_{imp}$ onto a Kondo problem \cite{PhysRev.149.491} that 
displays a resistivity minimum upon temperature lowering \cite{Hewson, Kondo}.  
Therefore an OV could be the magnetic impurity accounting for the observed resistivity upturn in two-dimensional SrTrO$_3$
under an applied electric field \cite{PhysRevLett.107.256601}.
We stress that independent of the coupling constant $g$, once $U$ is large enough (in this case $\gtrsim$ 1 eV),
the impurity level can at most trap only one electron. This is true even at high temperature. 
The nonzero impurity-bath coupling actually lowers
the impurity occupation, which can reduce the on-site repulsion,
and therefore allows for different charge states of the impurity inside the bulk gap \cite{PhysRevB.13.2553}.

Armed with this understanding, we now discuss some of the published calculations and experiments related to OV. 
First, since the correlation effect is important, a proper description of the ground state requires 
a many-body wave function (a wave function composed of more than one single Slater determinant),
and any DFT calculation  that by its nature replies on an independent-particle approximation, can only capture some parts of the physics. 
While using a hybrid functional gives an in-gap impurity level that explains the ARPES spectrum
\cite{PhysRevB.68.224105, PhysRevB.73.064106,  PhysRevB.86.155105}, 
calculations using LDA \cite{PhysRevB.70.214109, PhysRevB.86.155105, PhysRevB.86.195119} successfully give a metallic state. 
In many respects, the spin-polarized DFT using a GGA+U functional is the best option for this problem 
since it simultaneously explains both ARPES and transport measurements \cite{Hou_10}. 
However,  GGA+U gives a total magnetic moment.
Second, as it is the Coulomb repulsion that prevents doubly occupying the in-gap impurity level making OV the electron donor, the SrTiO$_3$ samples with OVs should be conductive at very low temperature, as indeed is observed in Refs. \cite{Kan_05, NatMetal_STO_2D}. 
Note that the strong repulsion originates from the OV-induced bound state being spatially localized, 
which is very different from an impurity state in doped semiconductors (P doped Si for example) whose effective Bohr radius is 
so large (typically 50 \AA $ $ or more)
that the Coulomb repulsion can be largely neglected \cite{Mermin}.
Third, as each OV at most captures one electron, and the nominal charge of an O atom is $2^-$, 
an OV produces an overall $attractive$ potential which can provide the interface  
confinement (quantum well) \cite{Santander-Syro_11,NatMetal_STO_2D,PhysRevB.86.195119,PhysRevB.86.125121}
and reduce the divergent potential of a polar interface (polar catastrophe) \cite{Hwang_06}.
Fourth, we notice that the OV-induced in-gap peak in ARPES is approximately 1.0 eV broad, Gaussian-like,
and has no significant temperature dependence between 20 and 300 K \cite{NatMetal_STO_2D,PhysRevB.86.195119}.
We attribute this large broadening to a difference in the local environment that leads to
a distribution of bonding state levels [via $\varepsilon_1$ and $t$ in Eq.~\eqref{eqn:H3o}], impurity-bath couplings [$g$ in
Eq.~\eqref{eqn:H_imp}], and therefore a distribution of in-gap peak positions.
Finally, we speculate that with the nonzero impurity-bath coupling (allowed only near the interface
or under the electric field), a single OV may account for the
Kondo resistance upturn observed in Ref. \cite{PhysRevLett.107.256601}. 
The localized electrons at different OV sites can interact via the conduction electrons,
and the Ruderman-Kittel-Kasuya-Yosida \cite{PhysRev.96.99,PhysRev.106.893}  or Zener 
double-exchange \cite{PhysRev.82.403} mechanism can be responsible for the interface ferromagnetism \cite{Brinkman_07, Li_11, Bert_11}.

In conclusion, the main message of this work is that an oxygen vacancy in SrTiO$_3$ should be regarded as a magnetic impurity,
whose deep level can only be partially filled due to its localized impurity wave function and therefore strong on-site repulsion.
Specifically, we show that an OV in SrTiO$_3$  simultaneously 
(1) results in a deep gap state, (2) donates one electron to the conduction band, and (3) provides
an attractive potential $without$ introducing a net magnetic 
moment to the system. These properties have been observed or implied by many experiments,
and can be easily understood by a three-orbital model.
Under the symmetry-reduced environment such as  the interface or under an electric field, 
the OV impurity level can couple to bulk bands, 
and an OV-based  Anderson impurity model is proposed.
This model can be used to study OV effects on the lightly n-doped SrTiO$_3$ including 
Kondo physics and interface ferromagnetism.

We thank Hosung Seo for the discussion of values of the Hubbard $U$, and
Richard Hatch and Agham Posadas for many helpful conversations.
Support for this work was provided through Scientific Discovery through Advanced Computing (SciDAC) program 
funded by U.S. Department of Energy, Office of Science, Advanced Scientific Computing Research and 
Basic Energy Sciences under award number DESC0008877.

\bibliography{OV_Impurity}

\begin{thebibliography}{56}
\bibitem[{cau()}]{caution1}
\bibinfo{note}{The $3d_{3z^2-r^2}$-based hybrid state is used to indicate that
  this state has a non-negligible Ti $4p_z$ component
  \cite{PhysRevB.86.161102}.}
\bibitem[{bat()}]{bath_orbital}
\bibinfo{note}{The correlation on the bath orbital can be estimated by
  evaluating the mean-field expectation value of the Hubbard interaction $U
  \sum_{i=1}^N n_{i,\uparrow} n_{i,\downarrow}$. For two electrons occupying
  the bath orbital, $\langle n_{i,\uparrow} \rangle = \langle n_{i,\downarrow}
  \rangle = 1/N$. This leads to an energy cost of $U \sum_{i=1}^N \langle
  n_{i,\uparrow} \rangle \langle n_{i,\downarrow} \rangle = U/N$, which
  approaches zero when the number of lattice sites $N \rightarrow \infty$.}
\bibitem[{mol()}]{molecular_U}
\bibinfo{note}{The on-site repulsion can be estimated by computing the
  expectation value $\langle A | U_1 \sum_{i=1,2} n_{i,\uparrow}
  n_{i,\downarrow}|A \rangle $ where $|A \rangle$ is the state doubly occupying
  the bonding level.}
\end{thebibliography}

\end{document}